# Political Bots and the Manipulation of Public Opinion in Venezuela

**Michelle Forelle**
USC Annenberg
forelle@usc.edu
@MCForelle

**Phil Howard**
University of Washington
pnhoward@uw.edu
@pnhoward

**Andrés Monroy-Hernández**
Microsoft Research, University of Washington
andresmh@uw.edu
@andresmh

**Saiph Savage**
UCSB, Universidad Nacional Autónoma México
saiph@cs.ucsb.edu
@saiphchen

**ABSTRACT**
Social and political bots have a small but strategic role in Venezuelan political conversations. These automated scripts generate content through social media platforms and then interact with people. In this preliminary study on the use of political bots in Venezuela, we analyze the tweeting, following and retweeting patterns for the accounts of prominent Venezuelan politicians and prominent Venezuelan bots. We find that bots generate a very small proportion of all the traffic about political life in Venezuela. Bots are used to retweet content from Venezuelan politicians but the effect is subtle in that less than 10 percent of all retweets come from bot-related platforms. Nonetheless, we find that the most active bots are those used by Venezuela's radical opposition. Bots are pretending to be political leaders, government agencies and political parties more than citizens. Finally, bots are promoting innocuous political events more than attacking opponents or spreading misinformation.

**FROM SOCIAL BOTS TO POLITICAL BOTS**

It is widely acknowledged that several regimes employ both people and bots to engage in political conversations online. The Chinese, Iranian, and Russian, governments employ their own social-media experts and pay small amounts of money to large numbers of people to generate pro-government messages.[1]

The word "botnet" comes from combining "robot" with "network," and it describes a collection of programs that communicate across multiple devices to perform some task. The tasks can be simple and annoying, like generating spam. The tasks can be aggressive and malicious, like choking off exchange points or launching denial-of-service attacks. Not all are developed to advance political causes. Some seem to have been developed for fun or to support criminal enterprises, but all share the property of deploying messages and replicating themselves.[2] There are two types of bots: legitimate and malicious. Legitimate bots, like the Carna Bot, which gave us our first real census of device networks, generate a large amount of benign tweets that deliver news or update feeds. Malicious bots, on the other hand, spread spam by delivering appealing text content with the link-directed malicious content.

Botnets are created for many reasons: spam, DDoS attacks, theft of confidential information, click fraud, cyber-sabotage, and cyber-warfare.[3] Many governments have been strengthening their cyberwarfare capabilities for both defensive and offensive purposes. In addition, political actors and governments worldwide have begun using bots to manipulate public opinion, choke off debate, and muddy political issues.

Social bots are particularly prevalent on Twitter.[4] They are computer-generated programs that post, tweet, or message of their own accord. Often bot profiles lack basic account information such as screen names or profile pictures. Such accounts have become known as "Twitter eggs" because the default profile picture on the social-media site is of an egg. While social-media users get access from front-end websites, bots get access to such websites directly through a mainline, code-to-code connection, mainly through the site's wide-open application programming interface (API), posting and parsing information in real time.

Bots are versatile, cheap to produce, and ever evolving. "These bots," argues Rob Dubbin, "whose DNA can be written in almost any modern programming language, live on cloud servers, which never go dark and grow cheaper by day."[5] Unscrupulous internet users now deploy bots beyond mundane commercial tasks like spamming or scraping sites like eBay for bargains. Bots are the primary applications used in carrying out distributed denial-of-service and virus attacks, email harvesting, and content theft. A subset of social bots are given overtly political tasks and the use of political bots varies across regime types.

**DATA, SOCIAL MEDIA, AND POLITICS IN LATIN AMERICA**

Twitter has become a powerful communication tool during many kinds of crises, political or otherwise. When the drug war erupted neither the drug lords nor the government expected a network of real-time war correspondents to spring up to report battles between police and gangs. Tweeting certainly didn't stop the drug war. But it helped people to cope. We can't measure how important the sense of online community provided by active tweeting can be in the first few weeks of a crisis, both in providing moral support and in keeping people safe. A few citizens rise to the occasion, curating content and helping to distinguish good information from bad.[2, p. 22], [6]

These are other examples of how people create their own public alert systems. When Hurricane Sandy hit Santiago, Cuba, information didn't come from the state; it came from the country's independent (illegal) journalists. Text messages about serious damage and the loss of life circulated among people a day before state media tried to bring citizens up to date.

However, governments and political leaders across Latin America are getting more adept at using digital media and managing data. And journalists in the region are getting better at covering data politics as a public issue. In March 2012 the Miami Herald reported that sensitive data about Venezuelans was being kept in Cuba.[7] Government databases, voting records, citizenship and intelligence records, and more were being stored in server farms outside Havana. To an outsider, it might be strange to think of Havana as a more secure city than Caracas, but of greater importance is the privacy issue of having data about a country's citizens was being transported out of country. Data-mining firms in Texas maintain detailed profiles of Argentina's citizens, and there is a global trade in data about people from all corners of the world.[8]

What is important in the Cuba-Venezuela connection is that the government did not choose to house important information with a firm or in a place that has good security or stable infrastructure. Data warehouses across the United States and Europe have such features. The network ties between Venezuela and Cuba are so strong that they overcame any technical logic to file storage. The data did not simply need to be stored, it needed to be stored with political compatriots that would share the same expectations of surveillance and social control. So while bad analysis of big data is a real danger, there is also a proven capacity for big data analysis and sophisticated bot use by authoritarian governments.

During recent political campaigns in Mexico, a political party paid close to $80,000 USD to a marketing company to create 22 Twitter trending topics in their favor, and a similar amount for promoting 576 tweets.[9] Bots tweeting from Venezuela were very active in the campaign by the Podemos party, a relatively newly established left-wing party in Spain, against their rival party Ciudadanos, a centrist party that has been dubbed the "Podemos of the right," during the recent national elections.[10]

While only a small number of Venezuelans use Twitter, the people who do use it tend to be urban, wealthy, young, and engaged in the political life of the country.[11] Around 14 percent of the active internet users in Venezuela also use Twitter—one of the highest Twitter adoption rates of any country. Furthermore, a recent study on global mobile trends from the Pew Research center reveals that more Venezuelans than any other people in Latin American share views about politics on social media. Venezuelans were the most likely to say that they learned someone's political beliefs were different than what they thought based on something they said in social media.[12]

Unfortunately, there is growing evidence that the country's social media conversations are being cluttered by political bots. A Freedom House report on the global state of freedom on the internet in 2013 acknowledged that "[r]ather than engaging in significant censorship, the [Venezuelan] government is making substantial use of social media platforms to propagate its point of view and counter political opposition".[13] A Twiplomacy study found that Nicolas Maduro is the third most effective world leader on Twitter (as measured by average number of retweets per tweet), but noted that it was odd that his tweets were favorited ten times less, a discrepancy that could suggest that bots are doing the majority of this retweeting.[14] Last summer, Cuban dissident Yusnaby Perez revealed that as many as 2,500 of the accounts retweeting President Maduro were bots.[15] In recent years, the Twitter accounts of opposition candidates, government critics and activists have been hijacked and used to disseminate pro-government messages.[13] This has gone both ways: in response to these hackings, there have been retaliation hacking campaigns by online activists against pro-government Twitter accounts.[13]

Other claims come directly from politicians. In early April of this year, President Maduro charged that Twitter in Venezuela was being managed (and manipulated) by his political enemies in the media.[16] In the past, Maduro has been quoted as saying "If lies come through Twitter we are going to strike back through Twitter."[13] In 2010, Diosdado Cabello, Speaker of the National Assembly and prominent member of the ruling party, was quoted saying that that "The opposition believes itself to be the owner of social networking," and claimed they would "assault social networks to counter the views expressed by [their] opponents."[17]

**SAMPLING AND METHOD**

The leading and opposition groups in Venezuela are active on Twitter, and politicians on both sides frequently use Twitter to address each other and their supporters. We identified six individual politicians who are particularly active Twitter users: four from the ruling United Socialist Party of Venezuela (PSUV), one from the opposition Democratic Unity roundtable (MUD), and one from the opposition Voluntad Popular (VP).

From the governing PSUV we tracked:

- **@NicolasMaduro:** Nicolás Maduro; 2.37 million followers. Currently the president of Venezuela, Maduro served under Hugo Chavez for several years as Minister of Foreign Affairs (2006-2013) and Vice President (2012-2013).

- **@dcabellor:** Diosdado Cabello; 1.24 million followers. Cabello, a powerful figure in the Maduro administration, is the Speaker of the National Assembly of Venezuela and widely recognized as one of the strong men of late president Hugo Chavez.

- **@TareckPSUV:** Tareck El Aissami; 849k followers. El Aissami is the governor of the state of Aragua, having served under Chávez as Interior and Justice Minister. Research conducted in Venezuela has suggested that the state government of Aragua has a particularly robust social media botnet strategy.

- **@luislopezPSUV:** Luis Lopez; 11k followers. Lopez is the Head of Health in the state of Aragua.

From the opposition we tracked:

- **@hcapriles:** Henrique Capriles Radonski; 5.23 million followers. Capriles is both the governor of the state of Miranda and the leader of the Democratic Unity roundtable (MUD), a coalition group of opposition parties. He ran as the opposition candidate against Hugo Chávez in 2012 and Nicolás Maduro in 2013, losing in 2013 by only a narrow (and hotly contested) margin.

- **@leopoldolopez:** Leopoldo López; 3.54 million followers. Former mayor of Chacao, and leader of the opposition group Voluntad Popular (VP), López was arrested in February 2014 for his involvement in the protests that occurred that month. Although he remains in jail, his Twitter account is now managed by his wife and supporters, and remains active.

While our purpose is to discern whether and how bots are being used to boost the messages of the ruling party, we included two accounts from the opposition in order to provide comparison points.

Our goal was to shed light on whether online bots worked for the Venezuelan government. We considered we might be able to detect possible bots by analyzing who is retweeting politicians content. We specifically examine the type of online platforms that people use to retweet as bots are known to use specific application to massively spread content.

Our research proceeded in three stages

- We collected all of the tweets generated by the main Venezuelan politicians.
- For each of their tweets we collected the list of people who had retweeted the content.
- We studied what platform each account used to retweet the content.

We also study the online profile of the people who retweeted the most content from politicians, particularly the individuals who used suspicious bot related platforms for their retweets.

Tracking bots—especially political bots—requires careful understanding of how the design features of platforms may constrain the sampling strategy. It is impossible to report the total number of bots engaged in Venezuelan politics. Twitter itself only allows researchers to get information from the 100 most recent retweets.[18] We assume that accounts using platforms like Botize or Masterfollow are bots because that is what those platforms are designed to do, and the accounts that use those platforms all retweet the same content at the same time. The Botize service advertises itself as a way to create your own bot and tasks after you have set up your own Twitter account.[19], [20] Rather trying to determine which specific tweet was generated by a bot we look at the type of platform used to retweet or create a tweet to determine the probability that message was bot-generated. Many of the accounts that are driven by a bot use bot-dominated platforms and normal platforms for tweeting activity. Indeed, many of the accounts that we identified as likely being bots have since been suspended by Twitter—the company also considered them to be bots.

We collected all of the tweets from the list of politicians for 2015 and the entire list of retweets for 2015. Note that Twitter API only allows for the collection of 100 latest retweets for a given tweet.

With these sampling caveats in mind, we captured and analyzed all of the tweets these key politicians generated between January 1$^{st}$ and May 31$^{st}$ 2015. We collected a total of 11,796 tweets. We then collected the retweet information of a subset of these tweets—who was retweeting, and platforms used to retweet. This process generated 205,077 retweets. Some 2 percent of these all of these retweets were bot generated. For some politicians as much as 5 percent of their retweet traffic comes from bots.

After taking an initial sample of tweets from our list of politicians, retweet information from the top 15 percent of most frequent retweets was gathered over the course of several days following each initial Tweet—amounting to 1,721 Tweets. Twitter also restricts how many queries researchers can make each hour so we took the most aggressive approach possible in collecting the retweet information of the most noteworthy tweets. The programming was done in python using the Tweepy API, which facilitates accessing and processing twitter data.

**FINDINGS AND ANALYSIS**

The flow of bot generated traffic may vary during political crises. But for our sample period, Table 1 presents the primary platforms for retweeting, and Table 2 details the percent of each accounts' retweets generated by bots. High profile accounts will always capture the attention of a small portion of bots, but with a closer look at the types of bots involved we can say more about how automated scripts are used to amplify political messaging—even when there is no political crisis.

**Table 1: Percent of Retweets by Venezuelan Politicians, by Platform, 2015**

| Platform | Percent |
|---|---|
| Twitter android | 34.3 |
| Twitter Webclient | 30.3 |
| Twitter iPhone | 14.9 |
| Twitter Blackberry | 11.9 |
| Tweetdeck | 2.3 |
| Botize | 1.7 |
| Mobileweb, | 0.9 |
| Twitter for Windows Phone | 0.8 |
| Twitter for Websites | 0.4 |
| Ubersocial | 0.3 |
| Masterfollow | 0.2 |

First, we mapped out the range of platforms being used to retweet the content generated by the sample of politicians.

Table 2: Percent of Retweets from Bot Platforms, By Politician, 2015

| Politician, Party, Account | Percent of Retweets Generated by Bots |
|---|---|
| Tareck El Aissami, Governing PSUV, @tareckpsuv | 0.4 |
| Henrique Capriles, Opposition DU, @hcapriles | 0.5 |
| Diosdado Cabello, Governing PSUV, @dcabellor | 0.7 |
| Luis Lopez, Governing PSUV, @luislopezpsuv | 1.9 |
| Nicolás Maduro, Governing PSUV, @nicolasmaduro | 2.0 |
| Leopoldo López, Opposition VP, @leopoldolopez | 4.4 |

We identified 11 platforms and found that approximately 38 percent of the retweets were generated by web browsers, and 62 percent were generated by mobile phone based apps. Of course, the majority of the politicians' content was retweeted by mobile phone applications, such as "Twitter for Android", "Twitter for Blackberry", or "Twitter for iPhone" because the country has one of the highest rates of mobile phone diffusion in Latin America.[21]

What is particularly fascinating is that people also used bot platforms, such as Botize or MasterFollow, to retweet. These are particular services define themselves has popular free services which can execute actions automatically. Altogether these kinds of platforms were used for less than 5 percent of all the retweets.

**Bots Used Mostly By Radical Opposition**
The use of bot platforms varied significantly by politician. For instance, Nicholas Maduro was the only politician who received retweets from the platform called *Retuitear al Presidente* (Retweet the President). Some politicians were also retweeted by bot platforms more than others. For instance, less than 1 percent of Diosdado Cabello's (@dcabellor) content was retweeted by bot platforms, whereas bot platforms altogether generated almost 7 percent of the retweets of tweets coming from the account of Leopoldo Lopez (@leopoldolopez).

Many of the bots that present themselves as official political candidates do so for specifically for a more radical opposition party (VP) rather than the dominant party (PSUV) or the more mainstream opposition party (MUD). In other countries it is the governing party that uses bots most aggressively, but in Venezuela the opposite is true. There has been relatively little bot activity, relative to the PSUV and VP, from the coalition of opposition Democratic Unity Party led by Capriles, the candidate who gave Maduro a serious run for his money in the 2013 election. At the same time, it is the VP whose leaders have been most aggressively targeted--and jailed—by the current government.

**More Bot Political Leaders than Bot Citizens**
We decided to profile the suspicious retweeters by inspecting the online Twitter profile of the top 10 percent most active retweeters who used a bot platform at least once. We identify two kinds of bot profiles.

*Bots Pretending to Be Government, Parties and Politicians.*
These bot accounts presented themselves as real political organizations, government offices, and real politicians, usually affiliated with the Voluntad Popular party. Many VP bot accounts represent themselves not as political candidates, but as VP party branches in different states and cities—they are branches of a party organization. For example, @VP_Carabobo is the account for the state of Carabobo, whereas @VPBejuma is the account for the small town of Bejuma in the state of Carabobo. Figure 2 is a screenshot of the @VPDiegoIbarra account page, which is an account that represents the municipality of Diego Ibarra in the state of Carabobo. All of these bots used the Botize platform to retweet content. In addition, most of their usernames had embedded in them the term "VP"--the initials of the Voluntad Popular party.

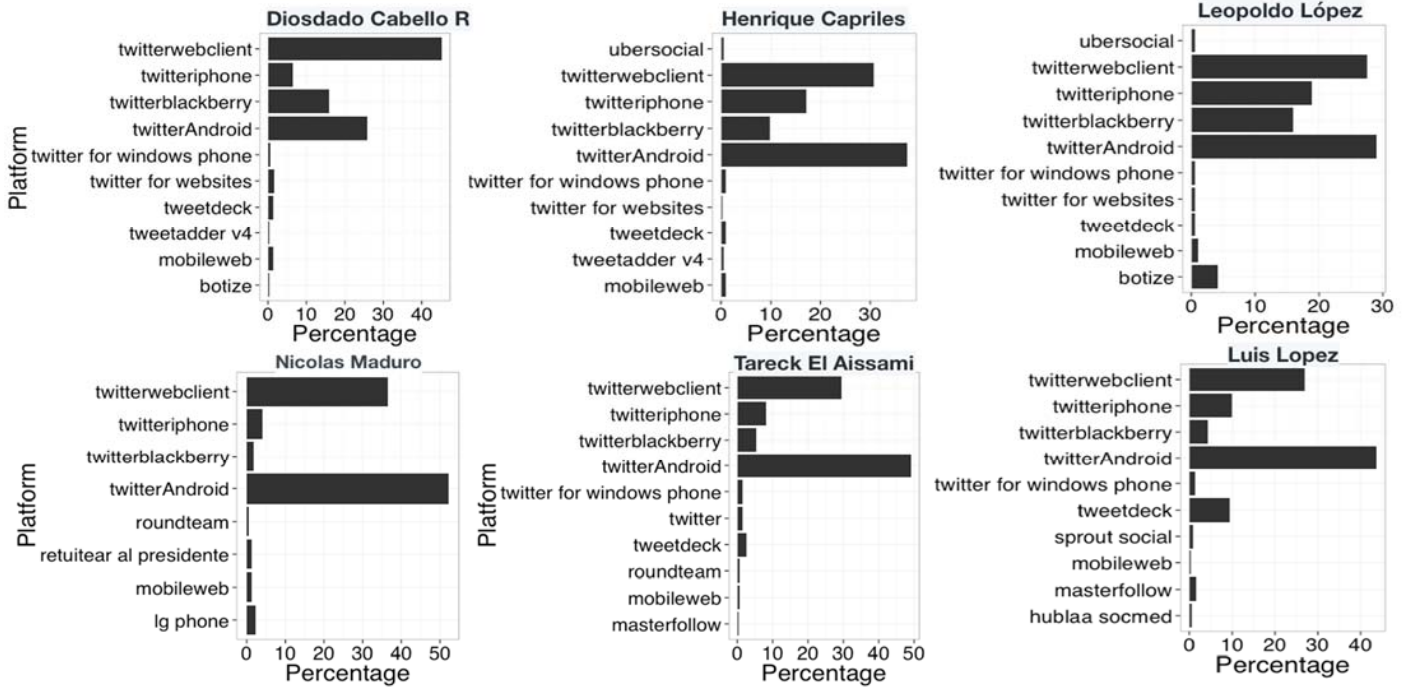

Figure 1: Platforms for Retweeting Venezuelan Politicians, Disaggregated by Politician, 2015

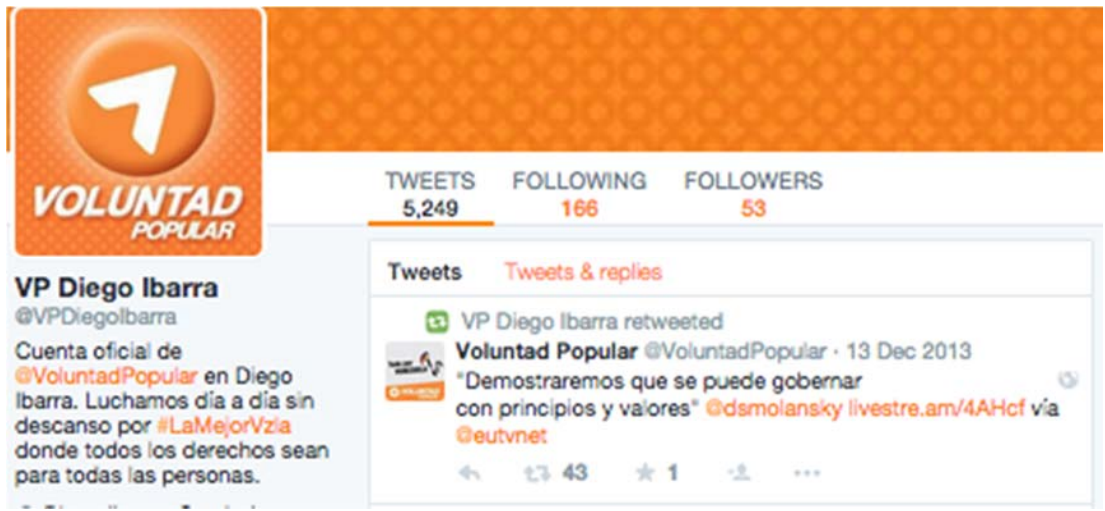

**Figure 2: Screenshot of Bot Account @VPDiegoIbarra**

For example, @VPBejuma has the consistent behavior of a bot-driven account. We believe the account exhibits suspicious bot-like behavior because it used the Botize interface to retweet over 100 tweets from other politicians. The account appears to only use Botize to retweet content, and Tweet Deck for its own tweets. Interestingly, we found that many such accounts use one particular bot platform, never multiple platforms, and that these bot accounts were only filled with tweets about Venezuelan politics. These bots do not retweet at the same time nor do they retweet the same content. But they primarily use bot related platforms for their retweets.

The account @VPBejuma reveals bot-like behavior more than @VPCarabobo because it almost exclusively retweets other notable VP accounts. And Figure 1 demonstrates that @VPDiegoIbarra has very similar tweeting behavior: thousands of tweets, following few other accounts, small number of followers. Yet this seems to be more a media strategy—creating localized accounts for VP branches on the part of the VP—rather than a bot planted by the PSUV to make the VP look bad. In contrast, @VPCarabobo tweets for a whole state, and thus might be more manually controlled than accounts for single municipalities such as @VPDiegoIbarra and @VPBejuma.

*Bots Pretending to be Citizens.*
We identified bots that were tasked with presenting themselves as somewhat average citizens. Each account identified a unique username, they didn't all include special patterns of initials as the bot politicians, and they didn't parrot the names of political parties or government offices. In addition, these citizen bots used a different set of platforms for retweeting. The primary platforms for bot citizens include TweetDeck (43 percent of bot citizens), MasterFollow (26 percent), and the main Twitter Web Client (25 percent). These Bot Citizens presented themselves very mechanically: many of them retweeted the exact same content at the exact same time. The content was exclusively related to Venezuelan politics. For example, Figure 3 offers additional screenshots: @salud_dia1 and @elchinito_pon retweeted the same content at the exact same time.

**Bots for Impression Management**
Bots have been used in many ways by political actors around the world. They have been used to attack opponents, choke off hashtags, and promote political platforms. During this sample period however we found that social media bots were used mostly for impression management. Social media bots are used for impression management in terms of a) spreading news about how leaders perform in public events within Venezuela and b) building the reputation that leaders are international statesmen in conversation with the leadership of other countries.

**CONCLUSIONS**
It is no secret that governments and political actors now make use of social robots or bots—automated scripts that produce content and mimic real users. Fake social media accounts now spread pro-governmental messages, beef up web site follower numbers, and cause artificial trends. Bot-generated propaganda and misdirection has become a worldwide political strategy.

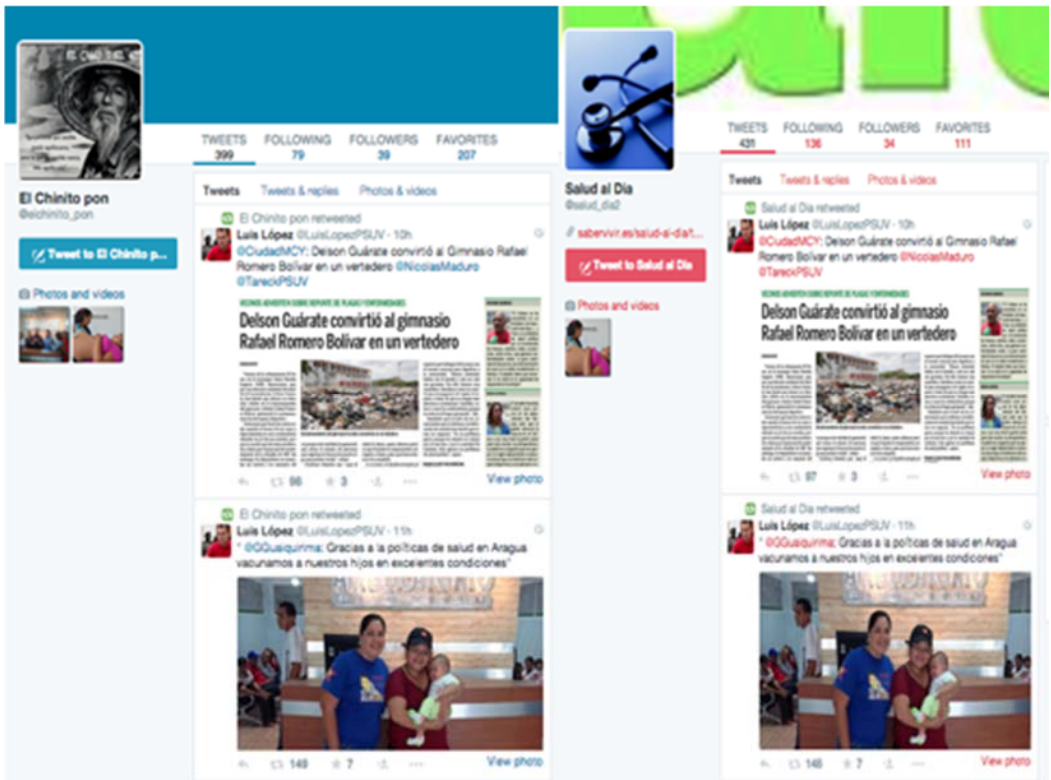

Figure 3: Screenshot of Bot Accounts @salud_dia1 and @elchinito_pon Retweet Incident

Robotic lobbying tactics have been deployed in several countries: Russia, Mexico, China, Australia, the United Kingdom, the United States, Azerbaijan, Iran, Bahrain, South Korea, Turkey, Saudi Arabia, and Morocco. Indeed, experts estimate that bot traffic now makes up over 60 percent of all traffic online—up nearly twenty percent from just two years ago.

Venezuelan government leaders also use bots to extend their social media impact. The use of bots in political conversation is particularly problematic given the number of real social media users incarcerated for using platforms like Twitter for political speech.[22] They have gone from simply padding follower lists to retweeting volumes of their own commentary and announcements. In this analysis, we find that bots generate a very small proportion of all the traffic about political life in Venezuela. Repeating this sample collection and study method in a time of political crisis or during an active election campaign would likely reveal different levels of bot prevalence and impact.

Bots are used to retweet content from Venezuelan politicians but the effect is subtle in that less than 10 percent of all retweets come from bot-related platforms. First, we find that the most active bots are those used by Venezuela's radical opposition. Second, bots are pretending to be political leaders, government agencies and political parties more than citizens. Finally, bots are promoting innocuous political events more than attacking opponents or spreading misinformation.

**ABOUT THE TEAM**

**Michelle C. Forelle**, a native of Venezuela, is a PhD student in Communication at the University of Southern California Annenberg School of Communication and Journalism. There, her research explores how emerging digital technologies are influencing how the courts and the public in America are interpreting such fundamental legal principles as property, privacy and free speech.

**Andrés Monroy-Hernández** is a researcher at Microsoft Research, and an affiliate faculty at the University of Washington. His work focuses on the design and study of social computing systems. Andrés was named one of the TR35 Innovators by the MIT Technology Review in Spanish, and one of CNET's influential Latinos in Tech. His research has received best paper awards at several computing conferences, recognized at Ars Electronica, and featured in The New York Times, The Guardian, NPR, and Wired. Andrés holds a Ph.D. from the MIT Media Lab, where he created the Scratch Online Community.

**Philip N. Howard** is a professor of communication, information and international affairs at the University of Washington and Central European University. Howard is the author of *The Managed Citizen* (Cambridge, 2006), the Digital Origins of Dictatorship and Democracy (Oxford, 2010), and most recently of *Pax Technica: How the Internet of Things May Set Us Free or Lock Us Up* (Yale, 2015). He is a frequent commentator on technology and politics for the national and international media. He blogs at www.philhoward.org and tweets from @pnhoward.

**Saiph Savage** is a computer science assistant professor at the University of West Virginia, and researcher at the National Autonomous University of Mexico (UNAM). Her research in human-computer interaction tackles problems in social computing, data visualization and crowdsourcing. The aim of her research group is to design social computing and crowdsourcing systems that spark collaborations and empower people to reach fulfilling goals. She is a Conacyt-UC MEXUS Doctoral Fellow, a Google Anita Borg Scholar, and a member of Microsoft's BizSpark. She holds a Ph.D. from the University of California, Santa Barbara

**ACKNOWLEDGEMENTS AND FUNDING DISCLOSURE**

The author(s) gratefully acknowledge the support of the National Science Foundation, "EAGER CNS: Computational Propaganda and the Production/Detection of Bots," BIGDATA-1450193, 2014-16, Philip N. Howard, Principal Investigator. Project activities were approved by the University of Washington Human Subjects Committee, approval #48103-EG. Any opinions, findings, and conclusions or recommendations expressed in this material are those of the author(s) and do not necessarily reflect the views of the National Science Foundation.

**REFERENCES**

[1] A. Chen, "The Agency," *The New York Times*, 02-Jun-2015.

[2] P. N. Howard, *Pax Technica: How the Internet of Things May Set Us Free or Lock Us Up*. New Haven: Yale University Press, 2015.

[3] S. Woolley and P. N. Howard, "Bad News Bots: How Civil Society Can Combat Automated Online Propaganda," *TechPresident*, 10-Dec-2014. .

[4] A. Samuel, "How Bots Took Over Twitter," *Harvard Business Review*, 19-Jun-2015. .

[5] R. Dubbin, "The Rise of Twitter Bots," *The New Yorker*.

[6] A. Monroy-Hernández, danah boyd, E. Kiciman, M. De Choudhury, and S. Counts, "The New War Correspondents: The Rise of Civic Media Curation in Urban Warfare," in *Proceedings of the 2013 Conference on Computer Supported Cooperative Work*, San Antonio, Texas, USA, 2013.

[7] J. O. Tamayo, "Fiber-optic Cable Benefiting Only Cuban Government," *The Miami Herald*, 25-May-2012.

[8] A. Tanner, "U.S.-Style Personal Data Gathering Is Spreading Worldwide," *Forbes*, 16-Oct-2013.

[9] J. Villamil, "El Precio de los Tuits Verdes," *NSS Tabasco: Información minuto a minuto*, 19-Jun-2015. .

[10] M. L. Congosto and A. Delgado, "Ni con Robots ni desde Venezuela: Así se Gestó la Campaña Tuitera Contra Ciudadanos," *El blog de el Español*, 05-Jun-2015. .

[11] "4 Ways How Twitter Can Keep Growing," *PeerReach Blog*, 07-Nov-2013. .

[12] "Emerging Nations Embrace Internet, Mobile Technology," *Pew Research Center's Global Attitudes Project*, 13-Feb-2014. .

[13] "Venezuela | Freedom on the Net 2013," Freedom House.

[14] "Twiplomacy Study 2015," Burson-Marsteller.

[15] Y. Perez, "¡Al Descubierto! Los Robots Retuiteadores de Nicolás Maduro (UPS)," *La Patilla*, 14-Jul-2014. .

[16] "Maduro Acusa a Hijo de Alberto Federico Ravell de 'Manejar' Twitter en Venezuela," *El Nacional*, 07-Apr-2015.

[17] M. Díaz Hernández, "'We Need to Be Careful Even of What We Think': Self-Censorship in Venezuela," *Global Voices*, 08-Feb-2015. .

[18] "How Can I Get All Retweets of a Specific Tweet?," *Twitter Developers*. [Online]. Available: https://twittercommunity.com/t/how-can-i-get-all-retweets-of-a-specific-tweet/11602. [Accessed: 17-Jul-2015].

[19] "CrunchBase | Botize," *CrunchBase*. [Online]. Available: https://www.crunchbase.com/organization/botize. [Accessed: 16-Jul-2015].

[20] "Botize," *Botize*. .

[21] "Venezuela: Internet Usage and Market Report," *Internet World Stats: Usage and Population Statistics*. [Online]. Available:

http://www.internetworldstats.com/sa/ve.htm. [Accessed: 04-Jul-2015].

[22] E. R. Biddle, M. Diaz, W. Li, H. Lim, and S. M. West, "Netizen Report: Leaked Documents Reveal Egregious Abuse of Power by Venezuela in Twitter Arrests," *Global Voices Advocacy*. .